\documentstyle[12pt,epsfig]{article}

\newlength{\dinwidth}
\newlength{\dinmargin}
\def\lapproxeq{\lower .7ex\hbox{$\;\stackrel{\textstyle
<}{\sim}\;$}}
\def\gapproxeq{\lower .7ex\hbox{$\;\stackrel{\textstyle
>}{\sim}\;$}}

\def\be{\begin{equation}}
\def\ee{\end{equation}}
\def\bea{\begin{eqnarray}}
\def\eea{\end{eqnarray}}

\begin{document}
\titlepage
\begin{flushright}
IPPP/01/31 \\
DCPT/01/62 \\
13 July 2001 \\
\end{flushright}

\vspace*{2cm}

\begin{center}
{\Large \bf Unintegrated generalised parton distributions} \\

\vspace*{1cm}
A.D. Martin$^a$ and M.G. Ryskin$^{a,b}$ \\

\vspace*{0.5cm}
$^a$ Department of Physics and Institute for Particle Physics Phenomenology, University of
Durham, Durham, DH1 3LE \\
$^b$ Petersburg Nuclear Physics Institute, Gatchina, St.~Petersburg, 188300, Russia
\end{center}

\vspace*{2cm}

\begin{abstract}
We show how the generalised (or skewed) parton distributions of the proton, $\overline{H} (x, \xi; k_t^2, \mu^2)$, 
unintegrated over the partonic transverse momenta, can be calculated from the known conventional parton 
distributions, $q (x, \mu^2)$ and $g (x, \mu^2)$, for small values of the skewedness parameter $\xi$.  
We demonstrate the procedure by numerically evaluating the skewed unintegrated gluon.  We also provide 
a simple approximate phenomenological form of the distribution, which may be used to make more 
rapid predictions of observables.
\end{abstract}

\newpage
\section{Introduction}

Generalised (or skewed) unintegrated parton distributions are
essential ingredients in the calculation of many high energy
processes initiated by protons.  Examples of such processes are
virtual photon Compton scattering $(\gamma^* p \rightarrow \gamma
p)$; diffractive heavy photon dissociation processes, like
diffractive vector particle electroproduction (e.g.\ $\gamma^* p
\rightarrow \rho p, J/\psi p, \Upsilon p, Zp$) or dijet production
$(\gamma^* p \rightarrow jj + p)$ or heavy quark production
$(\gamma^* p \rightarrow Q\bar{Q} + p)$; double diffractive
central particle production (e.g.\ Higgs or dijet production, $pp \rightarrow p + H + p,~p +
{\rm dijet} + p$).  In these examples, the $+$ sign is used to
indicate the presence of a rapidity gap.  

We use the `symmetrized' distributions introduced by Ji \cite{JI1}--\cite{JI3}
\begin{equation}
\label{eq:a1}
 \overline{H}_a (x, \xi; \mbox{\boldmath $k$}_t, \mbox{\boldmath $k$}_t^\prime;
 \mu^2),
\end{equation}
where $a = q, g$, with support $-1 \leq x \leq 1$.  They depend on the momentum fractions
$x_{1,2} = x \pm \frac{1}{2} \xi$ carried by the emitted and
absorbed partons, see Fig.~1.  Since, in general, $x_1 \neq x_2$,
we speak of skewed or {\it generalised} distributions.  We use the
term {\it unintegrated} to indicate that the transverse momenta
$\mbox{\boldmath $k$}_t, \mbox{\boldmath $k$}_t^\prime$ of the
partons have not been integrated over.  The bar on $\overline{H}$ is simply to denote that 
(\ref{eq:a1}) is the unintegrated distribution.  We reserve $H (x, \xi, \mu^2)$ for Ji's 
distributions, which are integrated over the transverse momenta.

\begin{figure}[h]
\begin{center}
\mbox{\epsfig{figure=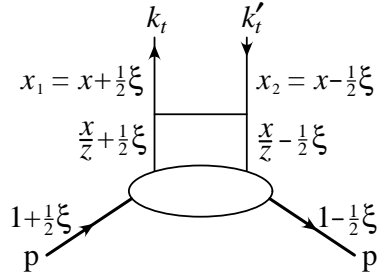,height=1.5in}}
\caption{Schematic diagram of the unintegrated generalised distribution $\overline{H} (x, \xi; 
 \mbox{\boldmath $k$}_t, \mbox{\boldmath $k$}_t^\prime; \mu^2)$ of (\ref{eq:a1}), showing the 
 last step of the evolution.  We use the `symmetrized' variables of Ji \cite{JI2}.}
\label{Fig1}
\end{center}
\end{figure}

The cross sections are calculated from these universal $\overline{H}$
distributions using the $k_t$ factorization approach, and the
explicit $k_t$ dependence allows the inclusion of the full
kinematic behaviour.  Assuming, for the moment, that $k_t$ has
been appropriately integrated over, then, in the $\xi \rightarrow
0$ limit, the distributions reduce to the conventional diagonal
quark and gluon distributions\footnote{There is a minor difference
with the distributions of Ji in that $H_g = xH_g^{\rm Ji}$.}
\begin{eqnarray}
 \label{eq:a2}
 H_q (x, 0) & = & \left \{ \begin{array}{ccl} q (x) & {\rm for} &
 x > 0 \\ - \bar{q} (-x) & {\rm for} & x < 0 \end{array} \right .
 \nonumber \\
 & & \\
 H_g (x, 0) & = & xg (x). \nonumber
\end{eqnarray}

The equations for the $\mu^2$ evolution of the distributions of
(\ref{eq:a1}) are well known, but as the $\overline{H}_a$ possess several
arguments it is impossible to determine the starting distributions
by fitting to the available sparse data.  However here we show
that it is possible to determine the $x, \xi, k_t^2, \mu^2$
dependence of the distributions from the values of the
conventional integrated distributions, $q (x, \mu^2)$ and $g (x,
\mu^2)$, that have been determined in the global analyses of deep
inelastic and related hard scattering data, assuming that $\xi^2 \ll
1$ and $\mbox{\boldmath $k$}_t^\prime \simeq \mbox{\boldmath
$k$}_t$. For most practical applications indeed $\xi^2 \ll 1$, since
it is specified by $\exp (- \Delta \eta)$ where $\Delta \eta$ is
the size of the rapidity gap. We also assume that $|t| \ll \mu^2$,
where $-t$ is the square of the momentum transfer. Indeed if $|t|$
were greater than the starting scale $\mu_0^2$, then the
logarithmic structure of the evolution from $\mu_0^2$ to $|t|$
would be destroyed.  Although we cannot predict the $t$ dependence, 
for small $t$ it is expected that the $t$ behaviour can be factored off as the 
proton form factor, see, for example, Ref.~\cite{JI3}.  

The procedure to determine generalised
unintegrated distributions is to carefully combine two existing
results. First, it has been shown that generalised (or skewed)
distributions are completely determined from knowledge of the
conventional distributions for small $\xi$ \cite{SGMR}. Second, it
has been found that the unintegrated distributions can be
determined from the conventional (integrated) distributions ---
the key observation is that angular ordering is only necessary in
the last step of the evolution \cite{KMR}.  

We briefly review these two results in Sections~2 and 3 below.  In Section~4 we 
proceed to use this information to derive a formula for the unintegrated generalised 
distributions.  Then in Section~5 we use the formula to compute the typical behaviour 
of the unintegrated skewed gluon distribution.  To gain insight we also show other 
predictions for the gluon.  Finally we give a simple phenomenological form that approximately 
reproduces the exact result.  Section~6 contains a brief conclusion.

\section{Generalised from conventional distributions}

Here we describe how the skewed (or generalised) distributions $H_a (x, \xi)$, with 
$a = q$ or $g$, can be determined from the conventional parton distributions, $q (x)$ or 
$g (x)$, for $\xi^2 \ll 1$.  For simplicity we omit the $\mu^2$ argument.  The essential result, 
due to Shuvaev \cite{SHUV}, is
\begin{equation}
\label{eq:a3}
 H_a (x, \xi) \; = \; \int_{-1}^1 \: dx^\prime \: K_a (x, \xi; x^\prime) \: f_a (x^\prime)
\end{equation}
with $a = q$ or $g$, where the kernels $K_a$ are known and where, for $\xi \ll 1,~f_a (x^\prime)$ reduces 
to the conventional distributions $q (x^\prime)$ or $g (x^\prime)$.  In this limit the Gegenbauer moments 
of the generalised distributions become equal to the Mellin moments of the usual parton distributions.  
For computational purposes, it is convenient to weaken the singularities in the integral expressions for 
$K_a$ by integration by parts.  Then, for $t = 0$, (\ref{eq:a3}) becomes\footnote{Note that here we 
adopt Ji's \cite{JI2} definition $\xi = (x_1 - x_2)$, whereas in Ref.~\cite{SGMR} $\xi \equiv 
(x_1 - x_2)/2$.} \cite{SGMR} 
\begin{eqnarray}
\label{eq:a4}
H_q (x, \xi) & = & \int_{-1}^1 \: dx^\prime \left [ \frac{2}{\pi} \: {\rm Im} \int_0^1 \frac{ds}{y (s) 
\sqrt{1 - y(s) x^\prime}} \right ] \: \frac{d}{dx^\prime} \left ( \frac{q (x^\prime)}{|x^\prime|} \right ), \\
& & \nonumber \\
\label{eq:a5}
H_g (x, \xi) & = & \int_{-1}^1 \: dx^\prime \left [ \frac{2}{\pi} \: {\rm Im} \int_0^1 \frac{ds \left (x + \xi 
\left (\frac{1}{2} - s \right ) \right )}{y (s) \sqrt{1 - y(s) x^\prime}} \right ] \: \frac{d}{dx^\prime} \left ( 
\frac{g (x^\prime)}{|x^\prime|} \right ),
\end{eqnarray}
where
\begin{equation}
\label{eq:a6}
y (s) \; = \; \frac{4 s (1 - s)}{x + \xi \left ( \frac{1}{2} - s \right )}.
\end{equation}

For small $x$, where we may assume the conventional distributions behave as
\begin{equation}
\label{eq:a7}
xf_a (x) \; \sim \; N_a \: x^{- \lambda_a},
\end{equation}
(\ref{eq:a4}) and (\ref{eq:a5}) may be greatly simplified.  Indeed we find that the ratios of the skewed to the diagonal 
distributions are 
\begin{equation}
\label{eq:a8}
\frac{H_a (x, \xi)}{H_a (x + \frac{1}{2} \xi, 0)} \; = \; R_a \left ( x/\xi, \lambda_a \right ),
\end{equation}
where $R_a$ are simple known functions \cite{SGMR}.

\section{Unintegrated from conventional distributions}

We outline the procedure \cite{KMR} which enables the distributions $f_a (x, k_t^2, \mu^2)$, unintegrated 
over $k_t$, to be calculated from the conventional (integrated) distributions $a (x, \mu^2) \equiv xq (x, 
\mu^2)$ or $xg (x, \mu^2)$.  We start from the DGLAP equation
\begin{equation}
\label{eq:a9}
\frac{\partial a (x, \mu^2)}{\partial \ln \mu^2} \; = \; \frac{\alpha_S}{2 \pi} \: \left [ \sum_{a^\prime} \: 
\int_x^{1 - \Delta} \: P_{aa^\prime} (z) \: a^\prime \left ( \frac{x}{z}, \mu^2 \right ) dz \: - \: a (x, \mu^2) \: 
V^a (\Delta) \right ],
\end{equation}
where in the virtual contribution
\begin{equation}
\label{eq:a10}
V^a (\Delta) \; = \; \sum_{a^\prime} \: \int_0^{1 - \Delta} \: P_{a^\prime a} (z) \: dz.
\end{equation}
In the case of the $g \rightarrow gg$ splitting we have to insert a factor $z$ in front of $P_{gg}$ in (\ref{eq:a10}), 
to account for the identity of the produced gluons.  The real emission term in (\ref{eq:a9}) gives the unintegrated 
parton density, except that we must allow for the modification due to virtual effects.  The virtual contribution 
does not change the parton $k_t$ and may be resummed to give the survival probability $T^a$ that parton $a$ remains 
untouched in the evolution up to $\mu^2$.  This survival factor is
\begin{equation}
\label{eq:a11}
T^a (k_t, \mu) \; = \; \exp \left (- \int_{k_t^2}^{\mu^2} \: \frac{dk_t^{\prime 2}}{k_t^{\prime 2}} \: 
\frac{\alpha_S (k_t^{\prime 2})}{2\pi} \: V^a (\Delta) \right ),
\end{equation}
and so the unintegrated distribution becomes
\begin{eqnarray}
\label{eq:a12}
f_a (x, k_t^2, \mu^2) & = & T^a (k_t, \mu) \: \left [ \frac{\partial a (x, k_t^2)}{\partial \ln k_t^2} 
\right ]_{\rm real} \nonumber \\
& & \\
& = & T^a (k_t, \mu) \left [ \frac{\alpha_S (k_t^2)}{2 \pi} \: \sum_{a^\prime} \: \int_x^{1 - \Delta} \: P_{a a^\prime} 
(z) \: a^\prime \left ( \frac{x}{z}, k_t^2 \right ) dz \right ]. \nonumber
\end{eqnarray}
It is at the last step of the evolution, shown in (\ref{eq:a12}), that the unintegrated gluon becomes dependent 
on $\mu^2$.  We have to make an appropriate choice of the cut-off $\Delta$ in (\ref{eq:a11}) and (\ref{eq:a12}).  
We impose angular ordering in the last step of the evolution \cite{CCFM,KMR}
\begin{equation}
\label{eq:a13}
\Theta (\theta - \theta^\prime) \; \Rightarrow \; \mu \: > \: \frac{zk_t}{1 - z}.
\end{equation}
Thus the maximum allowed value of the integration variable $z$ is
\begin{equation}
\label{eq:a14}
z_{\rm max} \; = \; \frac{\mu}{\mu + k_t}, \quad\quad\quad {\rm i.e.} \;\; \Delta \; \equiv \; 1 - z_{\rm max} \; 
= \; \frac{k_t}{\mu + k_t},
\end{equation}
and similarly $\Delta = k_t^\prime/(\mu + k_t^\prime)$ in (\ref{eq:a11}).

In summary, the $\mu$ dependence of the unintegrated distributions $f_a (x, k_t^2, \mu^2)$ enters at the last 
step, (\ref{eq:a12}), of the evolution through the survival factor $T^a$, which results from the resummation of the 
virtual contributions, and through the cut-off $\Delta$.

\section{Unintegrated generalised distributions}

To calculate the generalised (or skewed) {\it unintegrated} distributions $\overline{H}_a (x, \xi; k_t^2, \mu^2)$ from the 
conventional distributions $a (x, \mu^2)$, we combine the procedures outlined in Sections~2 and 3.  First, we 
extend (\ref{eq:a9}) so as to describe the evolution of the skewed ({\it integrated}) distributions $H_a (x, \xi, \mu^2)$
\begin{equation}
\label{eq:a15}
\frac{\partial H_a (x, \xi, \mu^2)}{\partial \ln \mu^2} \; = \; \frac{\alpha_S}{2 \pi} \left [ \sum_{a^\prime} 
\int_x^{1 - \bar{\Delta}} \: P_{aa^\prime} \left (z, \frac{\xi z}{x} \right ) \: H_{a^\prime} 
\left ( \frac{x}{z}, \xi, \mu^2 \right ) \: \frac{dz}{z} \: - H_a (x, \xi, \mu^2) (V_1^a + V_2^a) \right ]. 
\end{equation}
The splitting functions $P_{aa^\prime}$ are given by eq.~(22) of Ref.~\cite{JI2}, except that $P_{gg} 
(x, \xi) = x P_{gg}^{\rm Ji} (x, \xi)$, which arises because $H_g = xH_g^{\rm Ji}$.  For the 
antiquark contribution we should use $-H_q (-x/z, \xi, \mu^2)$ in (\ref{eq:a15}); see (\ref{eq:a2}).  
$V_1$ and $V_2$ are the virtual loop contributions corresponding, in the axial gauge, to the self-energies 
of the emitted and absorbed partons respectively,
\begin{equation}
\label{eq:a16}
V_i^a \; = \; \frac{1}{2} \: \sum_{a^\prime} \: \int_0^{1 - \Delta_i} \: P_{a^\prime a} (z) dz,
\end{equation}
in analogy to (\ref{eq:a10}).  We will specify the cut-offs on the $z$ integrations in a moment.

Resumming the virtual contributions, as before, introduces the survival factors $\sqrt{T_1}$ and 
$\sqrt{T_2}$, where now a square root occurs due to the factor $\frac{1}{2}$ in (\ref{eq:a16}).  In 
this way we obtain the $k_t^2$ dependence of the skewed distributions, namely
\begin{equation}
\label{eq:a17}
\overline{H}_a (x, \xi; k_t^2, \mu^2) \; = \; \sqrt{T_1^a T_2^a} \left [ \frac{\alpha_S (k_t^2)}{2\pi} \: 
\sum_{a^\prime} \: \int_x^{1 - \overline{\Delta}} \: \frac{dz}{z} \: P_{aa^\prime} \left (z, 
\frac{\xi z}{x} \right ) \: H_{a^\prime} \left (\frac{x}{z}, \xi, k_t^2 \right ) \right ]
\end{equation}
with
\begin{equation}
\label{eq:18}
T_i^a \; = \; \exp \left (- \int_{k_t^2}^{\mu^2} \: \frac{dk_t^{\prime 2}}{k_t^{\prime 2}} \: 
\frac{\alpha_S (k_t^{\prime 2})}{2 \pi} \: V_i^a (\Delta_i) \right ).
\end{equation}
Eq.~(\ref{eq:a17}) shows the modification necessary to extend (\ref{eq:a12}) to skewed distributions.  
As in Section~3, the cut-offs in the $z$ integrations of (\ref{eq:a17}) and (\ref{eq:a16}) are chosen 
so as to respect angular ordering in the last step of the evolution.  We obtain
\begin{eqnarray}
\label{eq:a19}
\overline{\Delta} & = & (1 - z_{\rm max}) \left (1 + \frac{\xi z_{\rm max}}{2x + \xi - \xi z_{\rm max}} 
\right ) \\
& & \nonumber \\
\label{eq:a20}
\Delta_1 & = & 1 - z_{\rm max} \\
& & \nonumber \\
\label{eq:a21}
\Delta_2 & = & (1 - z_{\rm max}) \left ( 1 + \frac{2 \xi z_{\rm max}}{2x + \xi - 2 \xi z_{\rm max}} \right ),
\end{eqnarray}
where $z_{\rm max} = \mu/\mu + k_t$ of (\ref{eq:a14}).  Although we have assumed that the emitted and 
absorbed partons of Fig.~1 have the same $k_t$, they have different momentum fractions, $x_1 \neq x_2$, 
and so the angular ordering gives different upper limits on the $z$ integration in (\ref{eq:a17}), and 
in (\ref{eq:a16}) for $V_1$ and $V_2$ \cite{MR}.

We are now in a position to use the results of Section~2.  We substitute (\ref{eq:a4}) and (\ref{eq:a5}) for the 
{\it integrated} skewed distributions $H_{a^\prime}$ into the right-hand-side of (\ref{eq:a17}).  In this way 
we determine the {\it unintegrated} skewed distribution $\overline{H}_a (x, \xi; k_t^2, \mu^2)$, 
for $\xi^2 \ll 1$, in terms of the conventional distributions, $q (x, \mu^2)$ or $g (x, \mu^2)$.

Note that there is no skewed effect coming from the $z \ll 1$ region of integration in (\ref{eq:a17}).  
In this domain the second argument $\xi z/x$ of the splitting function $P_{aa^\prime}$ is very small, 
while the first argument $(x/z \gg \xi)$ of the integrated distribution $H_{a^\prime}$ becomes large.  
As a consequence the $\xi$ dependence is negligible.  Therefore we may replace the singular $(1/z)$ 
part of $P_{aa^\prime}$ by the (diagonal) BFKL kernel, and use this approach even for partons 
which incorporate the BFKL contribution in the small $x$ region, just as was done in the diagonal 
case \cite{KMR}.  Recall that in Ref.~\cite{KMR} it was seen that the unintegrated partons calculated 
from pure DGLAP partons (as in Section~3) are very similar to those calculated using partons \cite{KMS} 
which incorporate BFKL effects.  It was concluded that the imposition of the angular-ordering constraint 
in the last step of the evolution is more important than the BFKL effects.

\section{Numerical results and discussion}

Typical behaviour of the unintegrated generalised gluon, $\overline{H}_g$, is shown by the continuous curves in 
Fig.~2.  The results are obtained by evaluating (\ref{eq:a17}) for $\overline{H}_g$, using MRST99 conventional 
parton distributions \cite{MRST99} to first determine the integrated distributions, $H_{a^\prime}$, from 
(\ref{eq:a4}) and (\ref{eq:a5}).  We show results for the most relevant case, $x_2 \ll x_1$, that is for 
$x = \xi/2$.  We take $\xi = 0.1, 0.01$ and 0.001 with $\mu = 10$~GeV.  To illustrate the $\mu$ dependence 
we also consider $\mu = 100$~GeV for the choice $\xi = 0.01$.  In contrast to DGLAP $k_t$ ordering, the angular 
ordering, (\ref{eq:a13}), imposed on the last step of evolution, enables the unintegrated distribution 
$\overline{H}_g$ to extend into the $k_t > \mu$ domain.  However, $\overline{H}_g$ decreases for large 
$k_t$ when $z_{\rm max}$ approaches $x$ --- the lower limit of integral (\ref{eq:a17}) (or (\ref{eq:a12})).

\vspace{-1.3cm}
\begin{figure}[!h]
\begin{center}
\mbox{\epsfig{figure=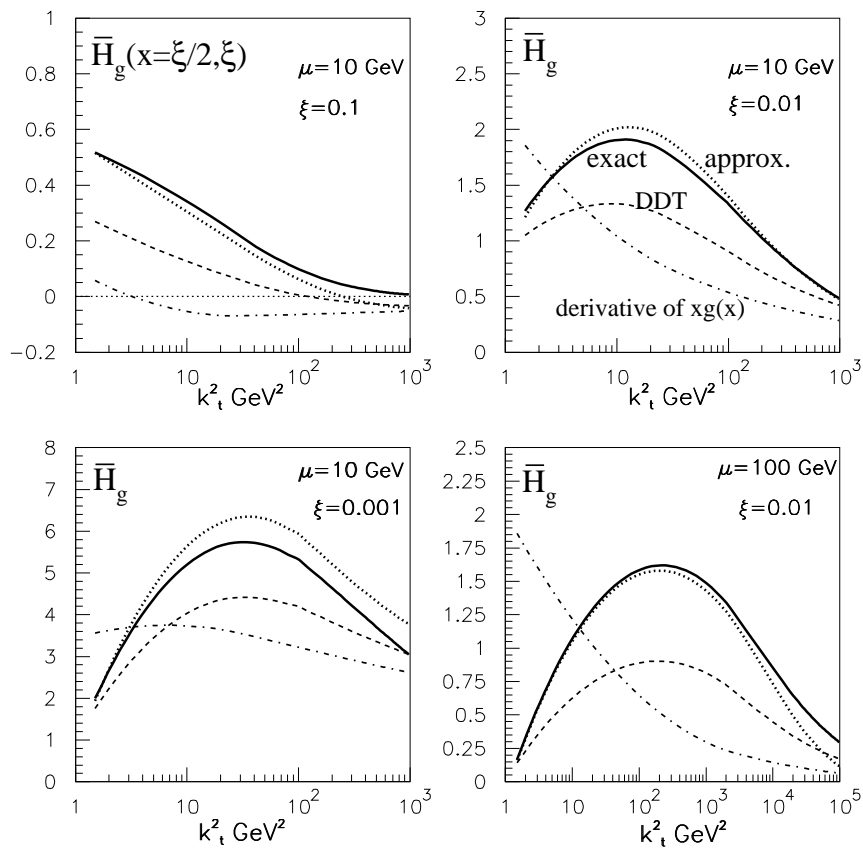,height=5in}}
\caption{The unintegrated skewed gluon distribution, $\overline{H}_g (x, \xi = 2x; k_t^2, \mu^2)$, as a 
function of $k_t^2$ for various values of $x$ and $\mu^2$.  The choice $\xi = 2x$ corresponds to the 
physically relevant fractions $x_1 = \xi$ and $x_2 = 0$ in Fig.~1.  The continuous curves are the full 
result obtained from MRST99 partons \cite{MRST99} by inserting (\ref{eq:a4}) and (\ref{eq:a5}) into 
(\ref{eq:a17}).  The dashed, dot-dashed and dotted curves are shown only for information and correspond, 
respectively, to the values obtained from (\ref{eq:a22}), (\ref{eq:a23}) and (\ref{eq:a26}).}
\label{Fig2}
\end{center}
\end{figure}

To gain insight, we compare the results with different approximate formula for $\overline{H}_g$.  The most naive 
possibility is to take
\begin{equation}
\label{eq:a22}
\overline{H}_g (x, \xi; k_t^2, \mu^2) \; = \; \left . \frac{\partial (y g (y, k_t^2))}{\partial \ln k_t^2} 
\right |_{y = x + \xi/2}.
\end{equation}
This simplified distribution, which is widely used in the literature, is shown by the dot-dashed curves in Fig.~2.  
For not too small $x$ and large $k_t^2$, formula (\ref{eq:a22}) gives a negative $\overline{H}_g$; the reason 
is that the virtual term in (\ref{eq:a9}) starts to dominate as $x$ increases.  This shortcoming may be avoided 
by developing a procedure originally implied in the work of DDT \cite{DDT}, see also \cite{KMR1}.  For 
generalised partons, this leads to the formula
\begin{equation}
\label{eq:a23}
\overline{H}_a (x, \xi; k_t^2, \mu^2) \; = \; \frac{\partial}{\partial \ln k_t^2} \left [ \sqrt{T_1^a T_2^a} \: 
a (y, k_t^2) \right ]_{y = x + \xi/2}.
\end{equation}
In the {\it diagonal} case $\xi = 0$ and $T_1^a = T_2^a = T^a$.  Then the derivative
\begin{equation}
\label{eq:a24}
\frac{\partial T^a}{\partial \ln k_t^2} \; = \; \frac{\alpha_S}{2 \pi} (V_1^a + V_2^a)
\end{equation}
cancels the virtual contribution in (\ref{eq:a9}), and $\overline{H}_a$ is positive everywhere \cite{KMR}.  
However for the {\it skewed} case the situation is worse.  Our sample curves correspond to $x = \xi/2$ for which 
$\Delta_2 = 1$ and $T_2^a = 1$.  Then only half of the negative term is cancelled by the derivative $\partial 
\sqrt{T_1^a}/\partial \ln k_t^2$, and even the DDT-improved formula (\ref{eq:a23}) may give negative 
$\overline{H}_a$; see the dashed curves in Fig.~2 for $\xi = 0.1$ and $k_t^2 > 100~{\rm GeV}^2$.  Note, 
however, the improvement of (\ref{eq:a23}) as compared to (\ref{eq:a22}) for the smaller $\xi$ values shown by 
Fig.~2.  Due to the presence of the $T_i^g$ survival factors in (\ref{eq:a23}), the gluon distribution $\overline{H}_g$ 
decreases for $k_t \ll \mu$ contrary to the behaviour of the simple derivative formula (\ref{eq:a22}).  This 
reflects the small chance for low $k_t$ partons to remain untouched in the long evolution up to $\mu$.

In order to reproduce the behaviour of the unintegrated generalised gluon $\overline{H}_g$, for $\xi^2 \ll 1$, 
in a simpler form than the full formula (\ref{eq:a17}), it is useful to modify the DDT-like approximation.  
We retain the well-justified and important leading logarithm contribution coming from 
\begin{equation}
\label{eq:a25}
\frac{\partial \sqrt{T^g}}{\partial \ln k_t^2} \; = \; \frac{N_C \alpha_S}{2 \pi} \: \ln \left (\frac{1}{1 - 
z_{\rm max}}\right ),
\end{equation}
but tune the non-leading contributions and an additional valence quark contribution (needed for $x \sim 0.1$) 
to approximately reproduce the full result shown by the continuous curves in Fig.~2.  The result is the
phenomenological form
\begin{eqnarray}
\label{eq:a26}
\overline{H}_g \left ( \frac{\xi}{2}, \xi; k_t^2, \mu^2 \right ) & = & \sqrt{T^g} \left [R_g \: \frac{\partial yg (y, 
k_t^2)}{\partial \ln k_t^2} \: + \: yg (y, k_t^2) \: \frac{N_C \alpha_S}{2 \pi} \left (\ln \frac{\mu + 
\frac{1}{2} k_t}{k_t} \: + \: 1.2 \frac{\mu^2}{\mu^2 + k_t^2} \right ) \right . \nonumber \\
& & \\
& & + \; 5 \left . \frac{\alpha_S}{2 \pi} \left (y u_{\rm val} (y, k_t^2) + yd_{\rm val} (y, k_t^2) \right ) 
\right ]_{y = \xi}, \nonumber
\end{eqnarray}
where $R_g$ is defined in (\ref{eq:a8}) and given in \cite{SGMR}.  Form (\ref{eq:a26}) gives the 
dotted curves in Fig.~2 and is a reasonable numerical approximation to the full result.  It is a rough and ready 
way to estimate the values of $\overline{H}_g$ for the physically relevant case $x = \xi/2$.

Of course, from a formal point of view, we can add to the generalised distribution, $\overline{H}$ of (\ref{eq:a17}), 
any function
\begin{equation}
\label{eq:a27}
\Delta \overline{H} (x, \xi) \; = \; f (x, \xi) \: \Theta \left (\frac{\xi}{2} - |x| \right )
\end{equation}
with support only in the time-like ERBL region, $|x| < \xi/2$ \cite{ERBL}.  Such a contribution is not obtained 
from the conventional partons determined from deep inelastic scattering (DIS) data.  However for the 
processes mentioned in the Introduction we only need the unintegrated skewed distributions in the space-like 
(DIS) domain.

\section{Conclusions}

We have presented a procedure which allows the unintegrated skewed parton distributions, $\overline{H}_a (x, \xi; 
k_t^2, \mu^2)$, to be calculated from the conventional distributions, $q (x, \mu^2)$ and $g (x, \mu^2)$, for 
$\xi^2 \ll 1$ and any $x$.  The key formula is (\ref{eq:a17}), with (\ref{eq:a4}) and (\ref{eq:a5}) as input.  
A crucial advantage of using unintegrated distributions is that they enable the full kinematics of the hard 
subprocess to be retained, and in this way incorporate a major part of higher order QCD corrections.  Moreover, 
the unintegrated skewed distributions are based on conventional partons fitted in global analyses of DIS and related 
data, and therefore have no uncertainties coming from the lack of knowledge of the (non-perturbative) initial 
conditions.  They can therefore be used to make predictions for physical processes, including those for 
diffractive production listed in the Introduction.

\section*{Acknowledgements}

This work was supported in part by the UK Particle Physics and Astronomy Research Council, the Russian 
Fund for Fundamental Research (grants 01-02-17095 and 00-15-96610), and the EU Framework TMR  
programme, contract FMRX-CT98-0194 (DG 12-MIHT).

\end{document}